\documentclass{article}

\usepackage{arxiv}

\usepackage[utf8]{inputenc} % allow utf-8 input
\usepackage[T1]{fontenc}    % use 8-bit T1 fonts
\usepackage{hyperref}       % hyperlinks
\usepackage{url}            % simple URL typesetting
\usepackage{booktabs}       % professional-quality tables
\usepackage{amsfonts}       % blackboard math symbols
\usepackage{nicefrac}       % compact symbols for 1/2, etc.
\usepackage{microtype}      % microtypography
\usepackage{lipsum}
\usepackage{graphicx}
\usepackage{multirow}
\usepackage{epstopdf}

\title{Experimental and theoretical calculation of gravity and moment of inertia using a physical pendulum}

\author{
  Alex Estupiñán\thanks{djlexes@gmail.com, personal email address}\\
  Facultad de Ciencias, Escuela de Física\\
  Universidad Industrial de Santander (UIS)\\
  Colombia, Santander, Bucaramanga \\
  \texttt{alex.estupinan@saber.uis.edu.co} \\
   \And
      Raul Ortiz \\
      Departamento De Matemáticas Y Ciencias Naturales\\
      Universidad Autónoma de Bucaramanga (UNAB)\\
      Colombia, Santander, Bucaramanga\\
      \texttt{rortiz579@unab.edu.co} \\
   \And
   Miguel Pico \\
   Universidad Autónoma de Bucaramanga (UNAB)\\
   Colombia, Santander, Bucaramanga\\
   \texttt{mpico602@unab.edu.co} \\
}

\begin{document}
\maketitle

\begin{abstract}
In the course of basic physics, more precisely the course of classical mechanics should be understood as clearly as possible the subject of rotational dynamics for students of science and engineering, to have clarity with the issues concerning rotational dynamics, such as calculation of torque and forces applied to a moving system. This paper presents the implementation of a physical pendulum for the physics laboratory using mainly a bar and a disc mounted on it, which can be moved along this bar, using implements such as a flexometer to measure the different lengths and a stopwatch to take the oscillation period of the pendulum.

This work shows the analytical development using the Simple Harmonic Motion (S.H.M) and experimental for the elaboration of the data collection and the realization of the laboratory with which the moment of inertia and the value of gravity could be obtained. Finally, the theoretical, experimental results and the respective errors obtained by the experiment are shown.
%\lipsum[1]
\end{abstract}

% keywords can be removed
\keywords{Rotational Dynamics \and Torque \and Simple Harmonic Motion (S.H.M).}

\section{Introduction}
One of the most interesting movements in physics is the rotational movement made by a physical pendulum, although the effort made to calculate the severity with this type of systems is very remarkable, it is very uncommon to find the detailed explanation of obtaining the calculations analytical using the study of a simple harmonic movement for this type of systems in which the rotational dynamics of the system can be studied.

In order to show the applicability and the importance of rotational movement, using an analytical model of linear behavior to be implemented in the physics laboratory, with this can to have a better understanding in this type of movement. Is for this that in this article the authors pretend to present a novel physics laboratory, in which can be show in detail the importance and also be able to validate a physical analytical and experimental model developmented of the authors, studying the dynamic behavior of the pendulum physics, for that used laboratory materials of easy access and low cost.

This paper is organized as follows:
Section \ref{analítico}, describes the analytical study corresponding to the calculates for the of the physical system. In Section \ref{experimental}, shown the experimental procedure for the data takes. In Section \ref{results} the results concerning to the validate of the analytic model using the experimentall method shown in the previous section.

\section{Analytical study}
\label{analítico}

\begin{figure}[h!]
	\centering
\includegraphics[width=0.5\textwidth]{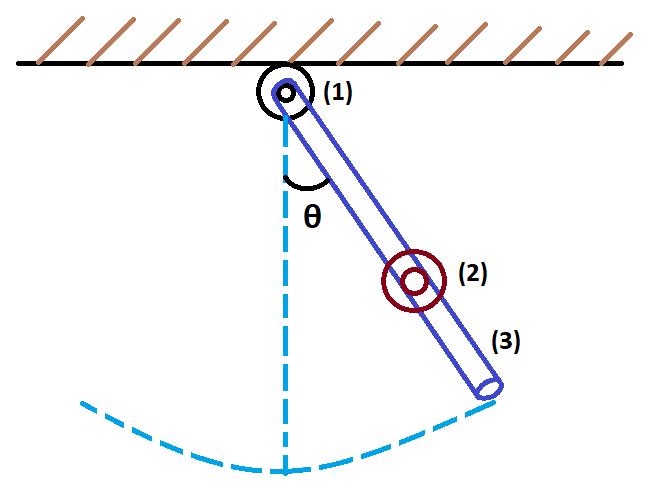}
	\caption{Physical system of pendulum to study.}
	\label{fig_1}
\end{figure}

The physical model of the system under study is shown in Figure \ref{fig_1}. The physical model of the system under study is shown in Figure 1. This system consists of (1) the point of support or pivot on which the system rests, (2) the disk which can be moved along the bar of iron (3).

To begin to analyze the dynamics of the movement to be studied, we must start from Newton's Second law for rotational dynamics, as follows:

\begin{equation}
	\sum_{i=0}^{n} \tau_{i} = I \alpha
\end{equation}

Carrying out the analysis from the point of support (1) of the Figure \ref{fig_1}, you can get the following equation \cite{sears}:

\begin{equation}
(-m_b g sin(\theta)) \frac{L}{2} + (-m_d g sin(\theta) L_x) = I \alpha,
\label{principal} 
\end{equation}

where $m_b$, $m_d$ are the mass of the bar and disk respectively. Expressing $I$ as the moment of total inertia of the system, being the sum of the moment of inertia of the bar $I_b$ plus that of the disk $I_d$.

\begin{equation}
 I = I_b + I_d.
 \label{momentos}
\end{equation}

The algebraic expressions for the moment of inertia of bar $I_b$ and disk $I_d$ using the Steiner's Theorem are \cite{serway}:

\begin{equation}
	I_b = \frac{1}{3} m_b L^2, \qquad
	I_d = m_d (L_y)^{2} + \frac{1}{2} m_d R^{2}
	\label{momentos_1}
\end{equation}

Replacing the expressions of Equation (\ref{momentos_1}) in Equation (\ref{momentos}), the moment of inertia of the system is obtained.

\begin{equation}
	I = \frac{1}{3} m_b L^2 + m_d (L_y)^{2} + \frac{1}{2} m_d R^{2},
	\label{momento_total}
\end{equation}

where $L$ are the lenght of bar, $L_y$ are the lenght of the position of the disk respect to the pivot and $R$ is the disk's radius.

Using the Equation (\ref{momento_total}) and replace this in the Equation (\ref{principal}), we are obtained:

\begin{equation}
(-m_b g sin(\theta)) \frac{L}{2} + (-m_d g sin(\theta) L_x) = \left( \frac{1}{3} m_b L^2 + m_d (L_y)^{2} + \frac{1}{2} m_d R^{2} \right) \alpha,
\end{equation}

by organizing the above equation, this can be written as:

\begin{equation}
g sin(\theta) \left[-m_b \frac{L}{2}-m_d L_y \right] = \left( \frac{1}{3} m_b L^2 + m_d (L_y)^{2} + \frac{1}{2} m_d R^{2} \right) \alpha,
\end{equation}

where $\alpha$  is the angular acceleration, if we writing this as $\ddot{\theta}$ and take this movement for small angles $sin(\theta) \approx \theta$, we can write the expression as follow: 

\begin{equation}
- \theta \left[ \left( m_b \frac{L}{2}+m_d L_y \right) \right] = \left( \frac{1}{3} m_b L^2 + m_d (L_y)^{2} + \frac{1}{2} m_d R^{2} \right) \ddot{\theta},
\end{equation}

Equaling to zero the before equation and comparing it with the second order differential equation corresponding to a simple harmonic oscillator \cite{zill}, we have:

\begin{equation} 
 \ddot{\theta} + \left( \frac{ g \left( m_b \frac{L}{2} + m_d L_y \right) } { \frac{1}{3} m_b L^2 + m_d (L_y)^{2} + \frac{1}{2} m_d R^{2} } \right) \theta = 0,  
\end{equation}

with,

\begin{equation}
	\ddot{\theta} + \omega^2 \theta=0.
\end{equation}

In this way it can be obtained that the expression for the angular frequency is given by the following expression:

\begin{equation}
	\omega^2 = \frac{ g \left( m_b \frac{L}{2} + m_d L_y \right) } { \frac{1}{3} m_b L^2 + m_d (L_y)^{2} + \frac{1}{2} m_d R^{2} },
\end{equation}

now taking into account the relationship of the angular frequency with the period of oscillation ($\omega = 2\pi/T$), the following expression can be obtained:

\begin{equation}
 \frac{4 \pi^2}{T^2} = \frac{ g \left( m_b \frac{L}{2} + m_d L_y \right) } { \frac{1}{3} m_b L^2 + m_d (L_y)^{2} + \frac{1}{2} m_d R^{2} }.
\end{equation}

Writing the above equation based on the square of the period $T^2$ we get to:

\begin{equation}
4 \pi^2 \left( \frac{1}{3} m_b L^2 + m_d (L_y)^{2} + \frac{1}{2} m_d R^{2} \right) = \left( m_b \frac{L}{2} + m_d L_y \right) g T^2,
\end{equation}

operating this last expression we have:

\begin{equation}
 \frac{4 \pi^2 m_d }{g} (L_y)^2 + \frac{4 \pi^2}{g} \left( \frac{1}{3} m_b L^2 + \frac{1}{2} m_d R^2   \right)  = m_b \frac{L}{2} T^{2} + m_d L_y T^{2} 
\label{analitical}
\end{equation}

\section{Experimental study}
\label{experimental}

In order to bring Equation (\ref{analitical}) to an algebraic expression of linear relationship, we can make the following substitution of physical variables:

\begin{equation}
b = \frac{4 \pi^2}{g} \left( \frac{1}{3} m_b L^2 + \frac{1}{2} m_d R^2   \right),
\label{constante_b}
\end{equation}

where the moment of inertia of the system measured from the center of mass of each object, both the bar and the disk $I_{CM}$ is \cite{tipler}:

\begin{equation}
 I_{CM} = \frac{1}{3} m_b L^2 + \frac{1}{2} m_d R^2,
 \label{momento_inercia_centro}  
\end{equation}

In addition we will call $\bar{m}$, $x$ and $y$ as:

\begin{equation}
\bar{m} = \frac{4 \pi^2 m_d }{g}, \quad x = (L_y)^2	\quad \mathit{and} \quad y = m_b \frac{L}{2} T^{2} + m_d L_y T^{2}.
\label{constantes} 
\end{equation}

Using the equation (\ref{analitical}), next we can write a linear dependence equation in the following way:

\begin{equation}
	y = \bar{m}x+b
	\label{model_lineal}
\end{equation}

The dimensions and masses of the objects used in the system to be studied are recorded in the Table \ref{table_1},

\begin{table}[h!]
	\centering
	\begin{tabular}{lllll}
		\cline{1-3}
		\multicolumn{1}{|l|}{\textbf{Object}} & \multicolumn{1}{l|}{\textbf{Length {[}m{]}}} & \multicolumn{1}{l|}{\textbf{Mass {[}kg{]}}} &  &  \\ \cline{1-3}
		\multicolumn{1}{|l|}{Bar}             & \multicolumn{1}{l|}{Long = 1}                & \multicolumn{1}{l|}{1.333}                  &  &  \\ \cline{1-3}
		\multicolumn{1}{|l|}{Disk}            & \multicolumn{1}{l|}{Radius= 0.05}            & \multicolumn{1}{l|}{1.611}                  &  &  \\ \cline{1-3}
		&                                              &                                             &  & 
	\end{tabular}
	\caption{Main measures of the physical pendulum.}
	\label{table_1}
\end{table} 

We perform the data collection in the following way, with the chronometer we record three times the time of 10 oscillations for the physical pendulum varying the length at which the disc is placed $L_y$, these data were recorded in the Table \ref{table_2},

\begin{table}[h!]
	\centering
	\begin{tabular}{|c|c|c|ll}
		\cline{1-3}
		\multicolumn{1}{|l|}{\textbf{Lenght $L_y$ {[}m{]}}} & \multicolumn{1}{l|}{\textbf{Time {[}s{]}}} & \multicolumn{1}{l|}{\textbf{Period  T {[}s{]}}} &  &  \\ \cline{1-3}
		\multirow{3}{*}{0.25}                               & 14.02                                      & \multirow{3}{*}{1.38}                           &  &  \\ \cline{2-2}
		& 13.79                                      &                                                 &  &  \\ \cline{2-2}
		& 13.78                                      &                                                 &  &  \\ \cline{1-3}
		\cline{1-3}
		\multirow{3}{*}{0.32}                               & 14.38                                      & \multirow{3}{*}{1.44}                           &  &  \\ \cline{2-2}
		& 14.73                                      &                                                 &  &  \\ \cline{2-2}
		& 14.24                                      &                                                 &  &  \\ \cline{1-3}
		\cline{1-3}
		\multirow{3}{*}{0.37}                               & 14.52                                      & \multirow{3}{*}{1.46}                           &  &  \\ \cline{2-2}
		& 14.64                                      &                                                 &  &  \\ \cline{2-2}
		& 14.71                                      &                                                 &  &  \\ \cline{1-3}
			\cline{1-3}
			\multirow{3}{*}{0.46}                               & 15.08                                      & \multirow{3}{*}{1.51}                           &  &  \\ \cline{2-2}
			& 15.06                                      &                                                 &  &  \\ \cline{2-2}
			& 15.10                                      &                                                 &  &  \\ \cline{1-3}
			\cline{1-3}
			\multirow{3}{*}{0.59}                               & 16.02                                      & \multirow{3}{*}{1.60}                           &  &  \\ \cline{2-2}
			& 16.13                                      &                                                 &  &  \\ \cline{2-2}
			& 16.04                                      &                                                 &  &  \\ \cline{1-3}
			\cline{1-3}
			\multirow{3}{*}{0.67}                               & 16.63                                      & \multirow{3}{*}{1.66}                           &  &  \\ \cline{2-2}
			& 16.64                                      &                                                 &  &  \\ \cline{2-2}
			& 16.48                                      &                                                 &  &  \\ \cline{1-3}
			\cline{1-3}
			\multirow{3}{*}{0.82}                               & 17.63                                      & \multirow{3}{*}{1.77}                           &  &  \\ \cline{2-2}
			& 17.75                                      &                                                 &  &  \\ \cline{2-2}
			& 17.72                                      &                                                 &  &  \\ \cline{1-3}
			\cline{1-3}
			\multirow{3}{*}{0.95}                               & 18.84                                      & \multirow{3}{*}{1.88}                           &  &  \\ \cline{2-2}
			& 18.76                                      &                                                 &  &  \\ \cline{2-2}
			& 18.80                                      &                                                 &  &  \\ \cline{1-3}
	\end{tabular}
	\vspace{0.4cm}
	
	\caption{Data capture for the period and length of the disk location $L_y$ in the experiment performed with the pendulum.}
	\label{table_2}
\end{table}

\section{Results}
\label{results}

To start analyzing the results that can be obtained using the analytical model of the Equation (\ref{model_lineal}) using the data the tables \ref{table_1} and \ref{table_2}, we are going to organize the values of $x$ and $y$ in the Table \ref{table_3}

\begin{table}[h!]
	\centering
	\begin{tabular}{ccllll}
		\cline{1-2}
		\multicolumn{1}{|c|}{\textbf{x {[}$m^2${]}}} & \multicolumn{1}{c|}{\textbf{y {[}$kg \cdot m \cdot s^{2}${]}}} &  &  &  &  \\ \cline{1-2}
		\multicolumn{1}{|c|}{0.0625}                 & \multicolumn{1}{c|}{2.0549}                                    &  &  &  &  \\ \cline{1-2}
		\multicolumn{1}{|c|}{0.1069}                 & \multicolumn{1}{c|}{2.4905}                                    &  &  &  &  \\ \cline{1-2}
		\multicolumn{1}{|c|}{0.1398}                 & \multicolumn{1}{c|}{2.7126}                                    &  &  &  &  \\ \cline{1-2}
		\multicolumn{1}{|c|}{0.2152}                 & \multicolumn{1}{c|}{3.2143}                                    &  &  &  &  \\ \cline{1-2}
		\multicolumn{1}{|c|}{0.3481}                 & \multicolumn{1}{c|}{4.1710}                                    &  &  &  &  \\ \cline{1-2}
		\multicolumn{1}{|c|}{0.4583}                 & \multicolumn{1}{c|}{4.8309}                                    &  &  &  &  \\ \cline{1-2}
		\multicolumn{1}{|c|}{0.6724}                 & \multicolumn{1}{c|}{6.2251}                                    &  &  &  &  \\ \cline{1-2}
		\multicolumn{1}{|c|}{0.9063}                 & \multicolumn{1}{c|}{7.7745}                                    &  &  &  &  \\ \cline{1-2}
		&                                                                &  &  &  & 
	\end{tabular}
	\caption{Data taken from the experiment for our linear model.}
	\label{table_3}
\end{table}

\begin{figure}[h!]
	\centering
	\includegraphics[width=0.7\textwidth]{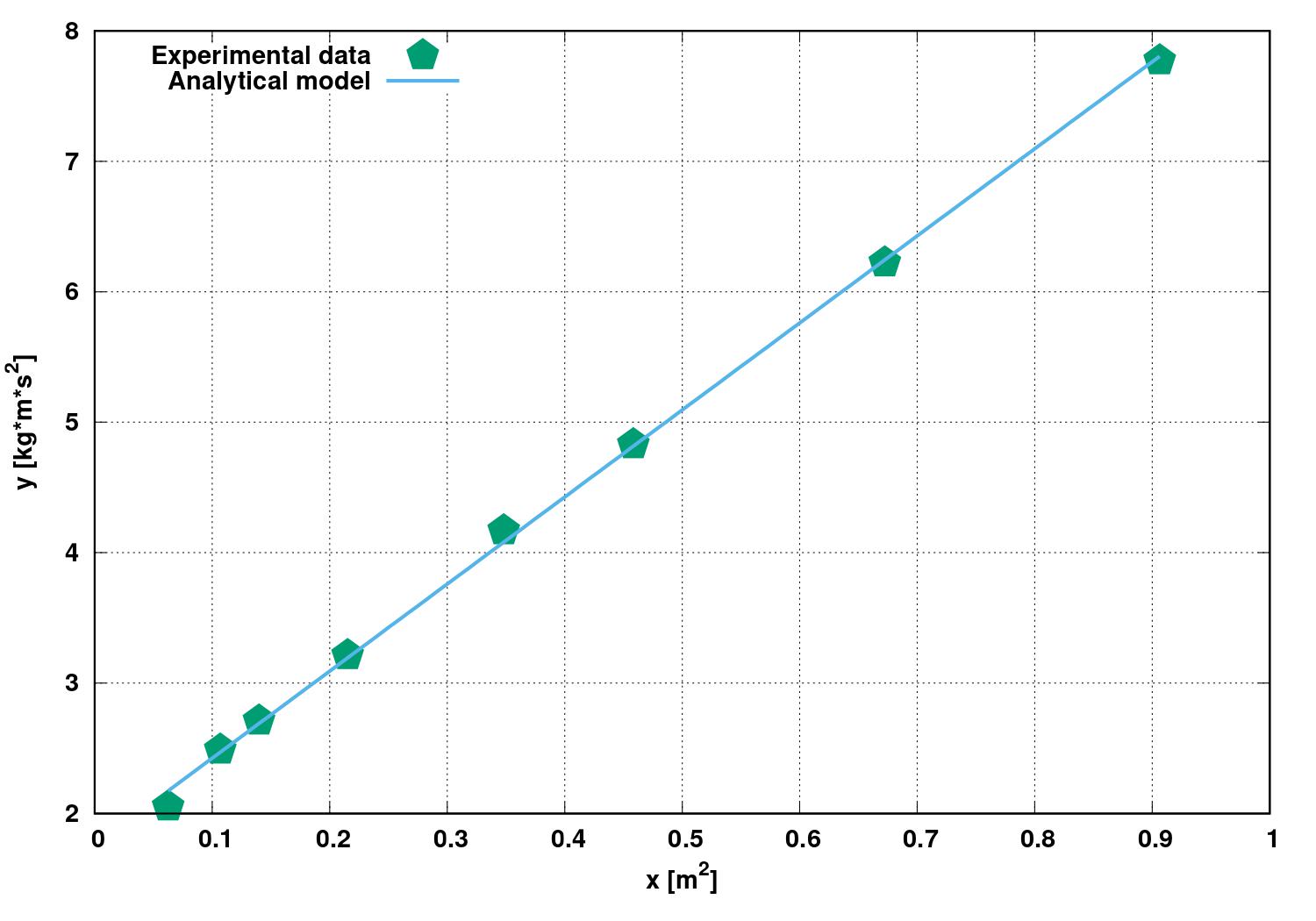}
	\caption{Experimental data and analytical model.}
	\label{plot_1}
\end{figure}

Realizing a lineal fitting to the experimental data \cite{gnuplot}, we can graph $x$ in function of $y$, we can compare the analytical model with the data taken in the experiment, this is shown in the Figure \ref{plot_1}.

The function obtained with the analytical linear (See the Equation (\ref{model_lineal})) adjustment was the following:

\begin{equation}
   y = 6.67381 \cdot x + 1.75689.
	\label{result}
\end{equation}

Using the value of the constant $\bar{m}$ obtained by Equation (\ref{constantes}) and compared with the Equation (\ref{result}), we can obtain the experimental value of gravity,

\begin{equation}
	g_{(exp)}= \frac{4 \pi^2 m_d}{\bar{m}} = 9.53 \quad m/s^2.
\end{equation}

Continuing to obtain the experimental moment of inertia with respect to the center of mass $I_{CM}$  of the system (bar and disk), it can be obtained using the equations (\ref{constante_b}), (\ref{momento_inercia_centro}) and (\ref{model_lineal}), in the following way:

\begin{equation}
	I_{CM(exp)}= \frac{b \cdot g_{(exp)}}{4 \pi^2} = 0.42 \quad kg \cdot m^2.
	\label{inertia_cm_exp}
\end{equation}

Using the theoretical results of local gravity $g$ and the moment of inertia $I_{CM}$ shown in the Table \ref{tabla_4}, one can calculate the experimental errors of gravity and moment of inertia.

\begin{table}[h!]
	\centering
	\begin{tabular}{|c|c|c|llllll}
		\cline{1-3}
	    & \textbf{Gravity }[m/$s^2$] & \textbf{Moment of inertia $I_{cm}$ {[}$kg \cdot m^2${]}} &  &  &  &  &  &  \\ \cline{1-3}
		\textbf{Theoretical value}  & 9.81                           & 0.44                                                     &  &  &  &  &  &  \\ \cline{1-3}
		\textbf{Experimental value} & 9.53                           & 0.42                                                     &  &  &  &  &  &  \\ \cline{1-3}
		\textbf{Absolute Error}     & 0.28                           & 0.02                                                     &  &  &  &  &  &  \\ \cline{1-3}
		\textbf{\% Relative Error}  & 2.85                           & 4.54                                                     &  &  &  &  &  &  \\ \cline{1-3}
	\end{tabular}
	\vspace{0.2cm}
	\caption{Experimental errors obtained from the analytical model.}
	\label{tabla_4}
\end{table}

\section{Conclusions}

In this work, it was possible to perform a new analytical approach and verify it through the realization of an experimental assembly, with the purpose of measuring the gravity and moment of inertia of a physical pendulum indirectly, using mainly simple materials. The results of this article showed very small errors, below 5\% which make it have a good reliability in the analytical model developed, in addition to the respective validation of this through the realization of the experiment using an innovative physical pendulum. In a future work we want to use this analytical approach to perform a numerical simulation of the rotational movement studied in this article.

\section*{Acknowledgments}

The authors would like to thank the Universidad Autónoma de Bucaramanga (UNAB), for lend us the installations and materials for carry to this experiment with which the analytical model presented in this paper could be validated.

\end{document}